\documentclass[12pt]{article}
\usepackage{graphicx}
\usepackage{lineno}

\newcommand\pubnumber{ATL-PHYS-PROC-2015-068}
\newcommand\pubdate{ }

\def\mcgill{McGill University, Rutherford Physics Building\\
Montr\'eal, Qu\'ebec\ \ H3A~2T8\\
CANADA}

\def\Title#1{\begin{center} {\Large #1 } \end{center}}
\def\Author#1{\begin{center}{ \sc #1} \end{center}}
\def\Address#1{\begin{center}{ \it #1} \end{center}}

\newcommand\pubblock{\rightline{\begin{tabular}{l} \pubnumber\\
         \pubdate  \end{tabular}}}
\newenvironment{Abstract}{\begin{quotation}  }{\end{quotation}}
\newenvironment{Presented}{\begin{quotation} \begin{center} 
             PRESENTED AT\end{center}\bigskip 
      \begin{center}\begin{large}}{\end{large}\end{center} \end{quotation}}
\def\Acknowledgements{\bigskip  \bigskip \begin{center} \begin{large}
             \bf ACKNOWLEDGEMENTS \end{large}\end{center}}

\def\beq{\begin{equation}}
\def\eeq#1{\label{#1}\end{equation}}
\def\eeqn{\end{equation}}

\def\beqa{\begin{eqnarray}}
\def\eeqa#1{\label{#1}\end{eqnarray}}
\def\eeqan{\end{eqnarray}}

\let\bar=\overbar

\def\Dslash{\not{\hbox{\kern-4pt $D$}}}
\def\dslash{\not{\hbox{\kern-2pt $\del$}}}

\def\msb{{\bar{\ssstyle M \kern -1pt S}}}

\begin{document}
\begin{titlepage}
\pubblock

\vfill
\Title{Measurements of $\alpha_{\mathrm{s}}$ in $pp$ Collisions at the LHC}
\vfill
\Author{ Andreas Warburton \\ (on behalf of the ATLAS and CMS Collaborations) }
\Address{\mcgill}
\vfill
\begin{Abstract}
The coupling of the strong force, $\alpha_{\mathrm{s}}$, is deemed to be a
fundamental parameter of Nature, and, beyond the quark masses,
constitutes the only free parameter in the QCD Lagrangian.  Provided
is an overview of CERN Large Hadron Collider (LHC) 
measurements of $\alpha_{\mathrm{s}}(M_Z)$ evaluated at the
$Z$-boson mass and of the running of $\alpha_{\mathrm{s}}(Q)$ as a function of
energy-momentum transfer $Q$. The measurements were performed by the
ATLAS and CMS Collaborations using proton-proton ($pp$) collisions 
with centre-of-mass energies of
7~TeV and data samples with time-integrated luminosities up to
5~fb$^{-1}$. Four different categories of observable were used in the
described extractions of $\alpha_{\mathrm{s}}$: inclusive jet cross sections,
3-jet to 2-jet inclusive cross-section ratios, 3-jet mass cross
sections, and top-quark pair production cross sections. These results,
which include the first NNLO measurement of $\alpha_{\mathrm{s}}$ at a hadron
collider and the first determinations of $\alpha_{\mathrm{s}}$ at energy scales
above 1~TeV, are consistent with each other, with the world-average
value, and with QCD predictions of their running with $Q$.
\end{Abstract}
\vfill
\begin{Presented}
Twelfth Conference on the Intersections of Particle and Nuclear Physics (CIPANP2015)\\
 $\phantom{}$ \\
Vail, Colorado, USA,  May 19--24, 2015
\end{Presented}
\vfill
\end{titlepage}
\def\thefootnote{\fnsymbol{footnote}}
\setcounter{footnote}{0}

\section{Introduction}

Due to the dominance of quantum chromodynamics (QCD) processes in its
collisions, the Large Hadron Collider (LHC)~\cite{Evans:2008zzb} is
foremost a factory of jet production.  Measurements of 
final states with jets at
the LHC are important to the testing of both the Standard
Model (SM) at high energies and perturbative QCD in
hitherto-unexplored kinematic regimes. Jets are frequently used in
characterizations of the principal backgrounds to searches for new
physics, as well as in the provision of constraints on parton
distribution functions (PDFs).

The coupling of the strong force, $\alpha_{\mathrm{s}}$, is deemed to be a
fundamental parameter of Nature, and, beyond the quark masses,
constitutes the only free parameter in the QCD
Lagrangian~\cite{Agashe:2014kda}. Experimentally testing the running
of the $\alpha_{\mathrm{s}}$ coupling with $Q$, the energy-momentum-transfer
scale, is important at frontier high energies, which have not yet been
studied. Moreover, investigations into $\alpha_{\mathrm{s}}$ can probe for new
physics by revealing deviations from the renormalization group
equation (RGE) calculation of the running of $\alpha_{\mathrm{s}}$ with
$Q$~\cite{Agashe:2014kda}.  For example, the existence of new coloured
matter near TeV energies could modify the three-jet rate, resulting in
significant departures from expectations in the apparent running of
$\alpha_{\mathrm{s}}$~\cite{Becciolini:2014lya}.

Leading-order dijet and inclusive jet production processes have
interaction rates proportional to $\alpha_{\mathrm{s}}^2$, whereas three-jet
production rates have $\alpha_{\mathrm{s}}^3$ dependencies.  A prominent strategy
in the experimental extraction of $\alpha_{\mathrm{s}}(Q)$ is to examine
inclusive jet cross-section {\it ratios} so as to eliminate or reduce
the effects of integrated-luminosity, PDF, and other systematic
uncertainties. Determinations of $\alpha_{\mathrm{s}}$ in LHC data can be
compared with the current world-average value of $\alpha_{\mathrm{s}}(M_Z) =
0.1185 \pm 0.0006$~\cite{Agashe:2014kda}, which was computed without
inputs from hadron-collider measurements. These proceedings constitute
a review of LHC Run~1 $\alpha_{\mathrm{s}}$ measurements using the
ATLAS~\cite{Aad:2008zzm} and CMS~\cite{Chatrchyan:2008aa} experiments
in studies of proton-proton ($pp$) collisions at centre-of-mass
energies of 7~TeV. Four different approaches to obtaining $\alpha_{\mathrm{s}}$
measurements will be discussed: inclusive jet cross sections, 3-jet to
2-jet inclusive cross-section ratios, 3-jet mass cross sections, and
top-quark pair production cross sections.

\section{Jet Measurements}

Both the ATLAS~\cite{Aad:2008zzm} and CMS~\cite{Chatrchyan:2008aa}
experiments use the infrared- and collinear-safe anti-$k_t$ jet
algorithm~\cite{Cacciari:2008gp}. ATLAS identifies jets by clustering
electromagnetic and hadronic calorimeter cells and using the values
0.4 and 0.6 for the anti-$k_t$ clustering parameter $R$. CMS forms
jets (with $R$ values of 0.5 and 0.7) by clustering particle-flow
candidates constructed by combining information from all the detector
subsystems.  Several PDFs are considered by the experiments; for
citations and discussions of the latest PDF sets, the reader is
referred to Ref.~\cite{Agashe:2014kda}.

A dominant experimental uncertainty is the degree of knowledge about
the jet energy scale (JES).  The ATLAS JES has been measured to have
uncertainties in the range $0.5 - 6$\% using {\it in situ} techniques
that correct for multiple $pp$ interactions~\cite{Aad:2014bia} and
depending on the jet transverse momentum ($p_{{\mathrm{T}}}$), the pseudorapidity,
and the event topology. The CMS JES has been determined
using various Monte Carlo and data techniques to have uncertainties
extrapolating from below 1\% in the vicinity of jet $p_{{\mathrm{T}}} \sim 150$~GeV
to 1.2\% near 2~TeV~\cite{CMS:2012zoa,Chatrchyan:2011ds}.

\section{Inclusive Jet Cross Sections}
\label{sec:inclusive}

An early analysis~\cite{Malaescu:2012ts} of inclusive jet cross
sections~\cite{Aad:2011fc} measured by ATLAS in 37~pb$^{-1}$ of 7~TeV
$pp$ collisions examined the running of $\alpha_{\mathrm{s}}(Q)$ up to $p_{{\mathrm{T}}} =
600$~GeV and made a measurement of $\alpha_{\mathrm{s}}(M_Z)$, the strong
coupling evaluated at the $Z$-boson mass.  No deviations from the RGE
running in QCD and the world-average value of $\alpha_{\mathrm{s}}(M_Z)$,
respectively, were observed. It is important to point out that the
obtained uncertainty on $\alpha_{\mathrm{s}}(M_Z)$ of 8\% was dominated by
differences between the results obtained using $R = 0.4$ and $R=0.6$
clustering parameters~\cite{Malaescu:2012ts}.  Understanding the
sensitivities of jet clustering sizes to nonperturbative QCD
corrections in contexts of NNLO (next-to-next-to-leading-order) QCD is
a critical prerequisite to enabling more precise extractions of
$\alpha_{\mathrm{s}}(M_Z)$ from inclusive jet cross sections at the LHC.

\begin{figure}
\centerline{
\mbox{\includegraphics[width=0.45\textwidth]{CMS-QCD-11-004_Figure_007-a.pdf}}
\mbox{\includegraphics[width=0.55\textwidth]{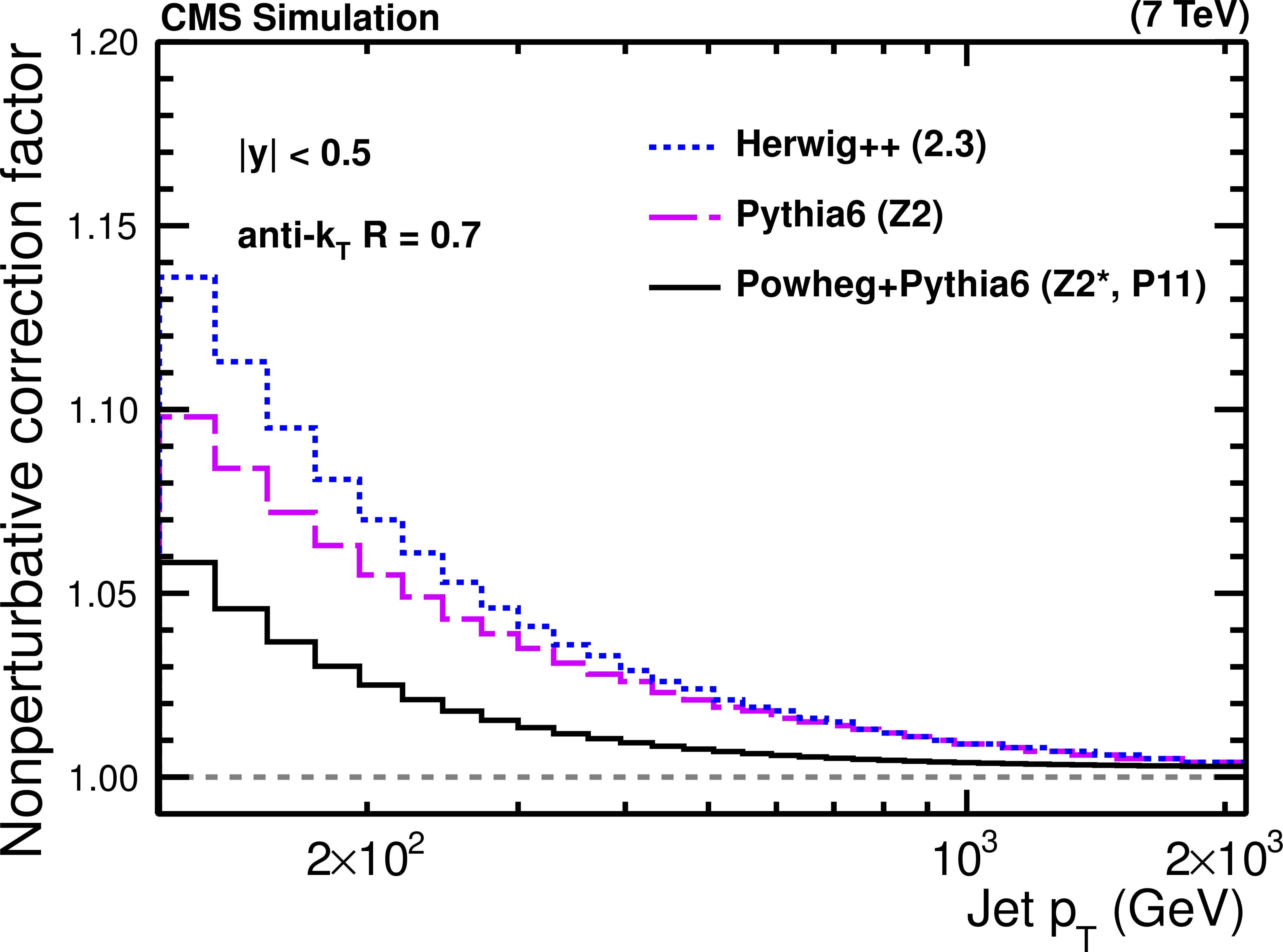}}
}
\caption{{\bf Left:} CMS inclusive jet cross sections for five
  different rapidity bins, for data (markers) and theory (thick lines)
  using the NNPDF2.1 PDF set~\cite{Chatrchyan:2012bja}. {\bf Right:}
  CMS nonperturbative corrections for the central regions in $|y|$ as
  derived in Ref.~\cite{Chatrchyan:2012bja}, using LO
  {\sc pythia6} tune Z2 and {\sc herwig++} with the default tune of
  version 2.3, in comparison with NLO corrections obtained from {\sc
    powheg} using {\sc pythia6} for showering with the two underlying
  event tunes P11 and Z2*~\cite{Khachatryan:2014waa}.}
\label{fig:cms_incl_xs}
\end{figure}

 Inclusive jet cross sections have also been measured in five rapidity
 ($y$) bins in 5.0~fb$^{-1}$ of 7~TeV LHC data by the CMS
 experiment~\cite{Chatrchyan:2012bja}. Refer to
 Fig.~\ref{fig:cms_incl_xs} (Left).  More recently, the CMS
 collaboration has used these jet cross sections to constrain PDFs and
 extract $\alpha_{\mathrm{s}}$~\cite{Khachatryan:2014waa}. The inclusive jet
 cross section, d$^2\sigma / {\rm d}p_{{\mathrm{T}}} {\rm d}y$, doubly differential in $p_{{\mathrm{T}}}$
 and rapidity $y$, is proportional to $\alpha^2_{\mathrm{s}}$. Perturbative QCD
 and reliable parton-shower Monte Carlo programs can be used to
 calculate this observable to next-to-leading order (NLO), but the
 extraction of $\alpha_{\mathrm{s}}$ first requires an estimate of the true
 theoretical differential cross section, taking into account
 nonperturbative corrections including multiple parton interactions
 and hadronization effects. Fig.~\ref{fig:cms_incl_xs} (Right) depicts
 the results of Monte Carlo calculations of these multiplicative
 nonperturbative corrections, as a function of jet $p_{{\mathrm{T}}}$, for
 determinations based on both leading-order (LO) and NLO matrix elements.

\begin{figure}
\centerline{
\mbox{\includegraphics[width=0.44\textwidth]{CMS-SMP-12-028_Figure_006-a.pdf}}
\mbox{\includegraphics[width=0.56\textwidth]{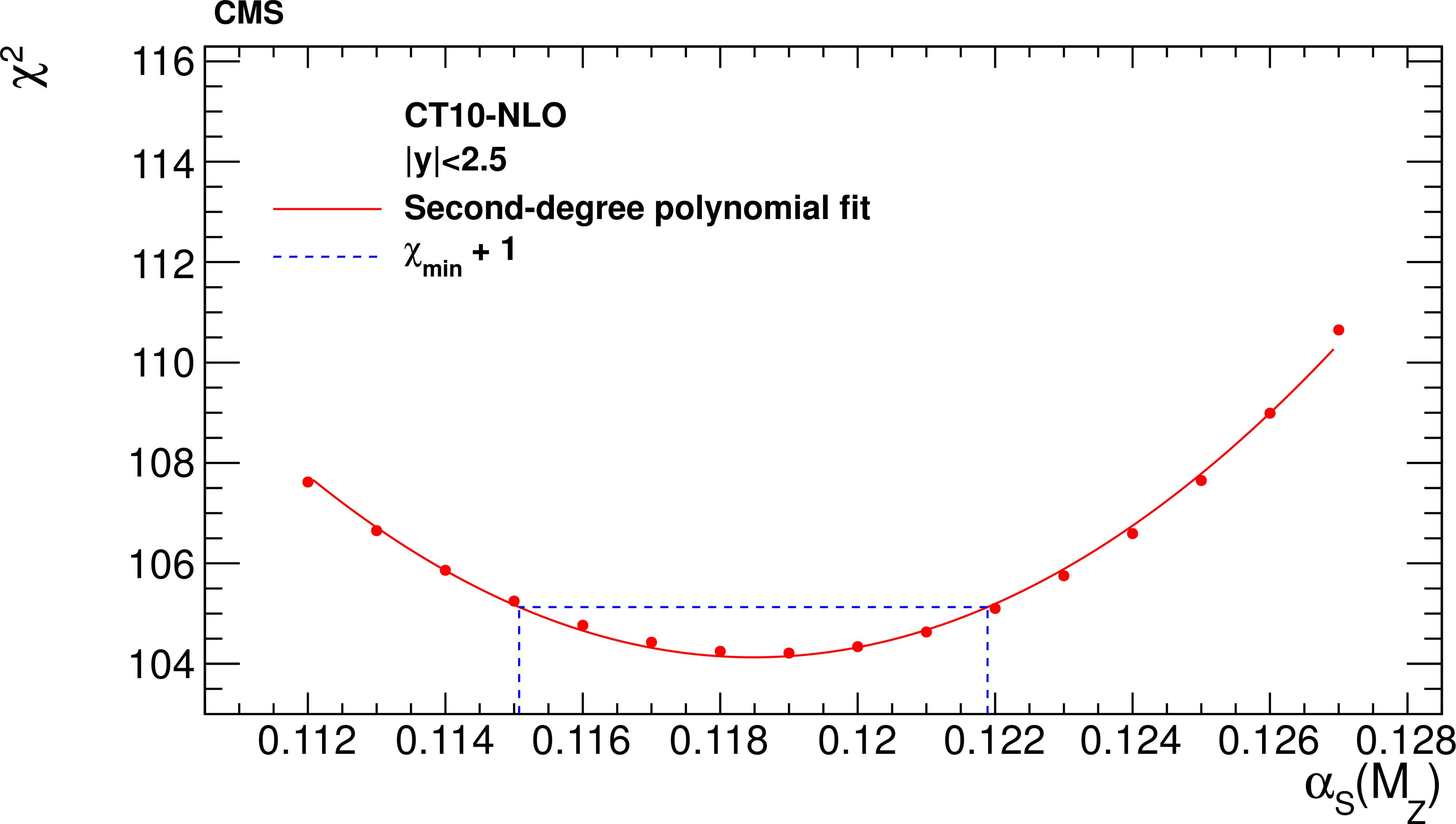}}
}
\caption{{\bf Left:} CMS ratio of inclusive jet cross section,
  corrected for nonperturbative (NP) and electroweak (EW) effects, to
  theoretical predictions using the CT10-NLO PDF set for the central
  rapidity bin, where the $\alpha_{\mathrm{s}}(M_Z)$ value is varied in the range
  0.112$-$0.126 in steps of 0.001. The error bars correspond to the
  total uncertainty~\cite{Khachatryan:2014waa}. {\bf Right:} CMS
  $\chi^2$ minimization with respect to $\alpha_{\mathrm{s}}(M_Z)$ using the
  CT10-NLO PDF set and data from all rapidity bins. The experimental
  uncertainty is obtained from the $\alpha_{\mathrm{s}}(M_Z)$ values for which
  $\chi^2$ is increased by one with respect to the minimum value,
  indicated by the {\it dashed line}. The {\it curve} corresponds to a
  second-degree polynomial fit through the available $\chi^2$
  points.~\cite{Khachatryan:2014waa}.}
\label{fig:cms_ratio_chisq}
\end{figure}

The extraction of $\alpha_{\mathrm{s}}$ from inclusive jet cross-section data is
achieved by forming ratios of the data cross-section observations,
corrected for nonperturbative (NP) and electroweak (EW) effects, with
theoretical predictions computed using different input values of
$\alpha_{\mathrm{s}}$ and various PDF sets.  An example of a set of predictions
with $\alpha_{\mathrm{s}}$ ranging from 0.112 to 0.126 in steps of 0.001 and
using the CT10-NLO PDF set is provided in
Fig.~\ref{fig:cms_ratio_chisq} (Left).  A $\chi^2$ is formed between
the measurements and the theoretical predictions (refer to
Fig.~\ref{fig:cms_ratio_chisq} (Right)), where the covariance matrix
entering into the $\chi^2$ definition includes terms that are statistical (with
correlations), uncorrelated systematics, JES, unfolding, luminosity,
and PDF related.  The $\chi^2$ is minimized and the extracted result,
with all rapidity bins $|y|<2.5$ combined,
is~\cite{Khachatryan:2014waa}
\begin{equation}
\alpha_{\mathrm{s}}(M_Z) = 0.1185\pm 0.0019({\rm exp})\pm 0.0028({\rm PDF})\pm 0.0004({\rm NP})
 \phantom{}^{+0.0053}_{-0.0024}({\rm scale}),
\end{equation}
where the uncertainties are experimental, PDF, nonperturbative, and
scale-related, respectively.  The total uncertainty varies between
3.5\% and 5.5\%, and is dominated by the renormalization ($\mu_r$) and
factorization ($\mu_f$) scales, the uncertainties for which are
computed by the standard method of varying the default values, chosen
in this instance to be $\mu_r = \mu_f = p_{{\mathrm{T}}}$, by factors of 0.5 and 2.
The $\chi^2$ minimization fit is also performed in separate $p_{{\mathrm{T}}}$ bins
in order to test the running of $\alpha_{\mathrm{s}}$ as a function of
$Q$~\cite{Khachatryan:2014waa}.

\section{3-Jet to 2-Jet Inclusive Cross-Section Ratios}

\begin{figure}
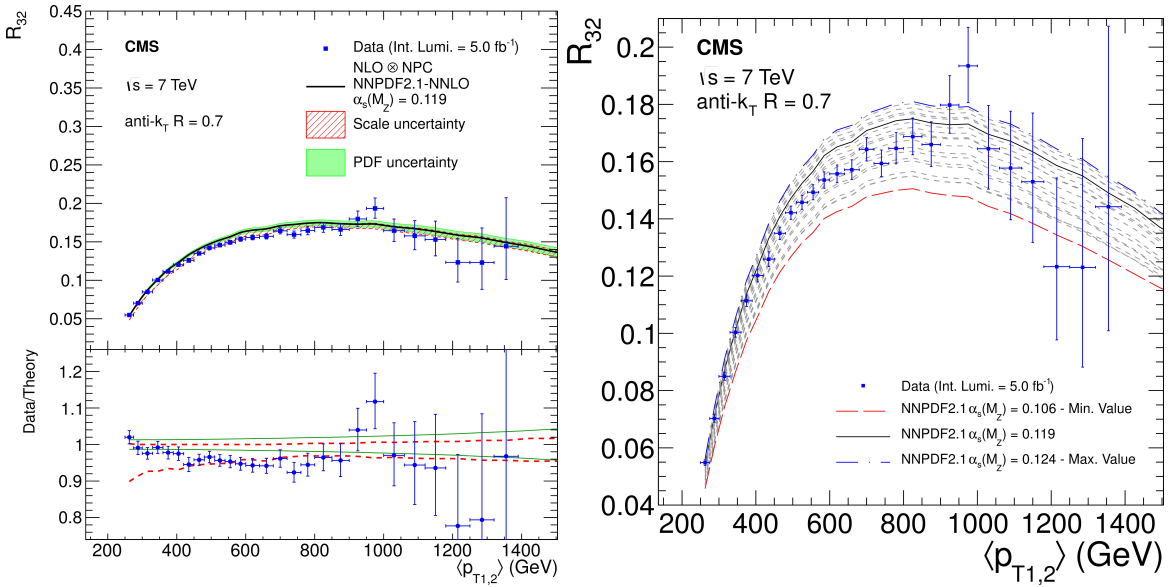

\centerline{
\mbox{\includegraphics[width=0.48\textwidth]{CMS-QCD-11-003_Figure_001-a.pdf}}
\mbox{\includegraphics[width=0.52\textwidth]{CMS-QCD-11-003_Figure_002-a.pdf}}
}
\caption{{\bf Left:} CMS measurement of $R_{32}$ and
  next-to-leading-order predictions using the NNPDF2.1 PDF set. {\bf
    Right:} Next-to-leading-order predictions using the NNPDF2.1 PDF
  set and compared with the CMS $R_{32}$ measurements for several
  assumed values of $\alpha_{\mathrm{s}}(M_Z)$.  Results using other PDF sets are
  also provided in Ref.~\cite{Chatrchyan:2013txa}.}
\label{fig:cms_R32}
\end{figure}

The observable $R_{32}$, defined as the ratio of the inclusive 3-jet
cross section to the inclusive 2-jet cross section, is proportional to
$\alpha_{\mathrm{s}}(Q)$.  Here, $Q$ is defined as the average transverse
momentum of the two highest-$p_{{\mathrm{T}}}$ (leading) jets in a selected event:
$Q \equiv \langle p_{{\mathrm T}_{1,2}}\rangle = \frac{p_{{\mathrm T}_1} + p_{{\mathrm T}_2}}{2}$.
Figure~\ref{fig:cms_R32} (Left) depicts a CMS measurement of $R_{32}$
versus $\langle p_{{\mathrm T}_{1,2}}\rangle$ for selected events with two or
more jets having $p_{{\mathrm{T}}} > 150$~GeV and rapidities $|y_{1,2}| <
2.5$~\cite{Chatrchyan:2013txa}.

Figure~\ref{fig:cms_R32} (Right) represents a comparison of the CMS
observed $R_{32}$ distribution with numerous NLO predictions in which
the assumed values of $\alpha_{\mathrm{s}}(M_Z)$ were varied in steps of 0.001
from 0.106 to 0.124.  Using techniques similar to those described in
Section~\ref{sec:inclusive}, fits were performed to identify the value
of $\alpha_{\mathrm{s}}$ most favoured in the data~\cite{Chatrchyan:2013txa}:
\begin{equation}
\alpha_{\mathrm{s}}(M_Z) = 0.1148\pm 0.0014({\rm exp})\pm 0.0018({\rm PDF})\pm 0.0050({\rm theory}),
\end{equation}
with a total uncertainty of 4.7\% that was again dominated by
sensitivity to knowledge about the impact of the $\mu_r$ and $\mu_f$
scales, defined identically to those discussed in
Section~\ref{sec:inclusive}. The analysis was also done splitting the
sample into three different $p_{{\mathrm{T}}}$ ranges, thereby providing running
values of $\alpha_{\mathrm{s}}(Q)$ of $0.0936\pm 0.0041$, $0.0894\pm 0.0031$, and
$0.0889\pm 0.0034$ for $Q = 474$, 664, and 896~GeV,
respectively~\cite{Chatrchyan:2013txa}.

In 37~pb$^{-1}$ of early 7~TeV data, the ATLAS collaboration also
performed a preliminary study of $R_{32}$, but as a function of the
$p_{{\mathrm{T}}}$ of the leading jet~\cite{ATLAS:2013lla}.  In addition, an
observable $N_{3/2}$, defined as the ratio of the jet $p_{{\mathrm{T}}}$
distribution for events with at least three jets to that for events
with at least two jets, was used as a function of jet $p_{{\mathrm{T}}}$. In
contrast to $R_{32}$, for which the numerator and denominator received
a single value per event, the numerator and denominator used to
determine $N_{3/2}$ contained one entry per jet. The $N_{3/2}$
approach was found by the ATLAS collaboration to have less sensitivity
to the JES and to event pile-up effects.  The $\alpha_{\mathrm{s}}$ values from
this study~\cite{ATLAS:2013lla} were consistent with the world
average~\cite{Agashe:2014kda} and possessed total uncertainties
ranging from 6\% to 15\%, dominated by the effects of uncertainties
due to the renormalization and factorization scale parameters, $\mu_r
= \mu_f = p_{{\mathrm{T}}}$.

\section{3-Jet Mass Cross Sections}

\begin{figure}
\centerline{
\mbox{\includegraphics[width=0.5\textwidth]{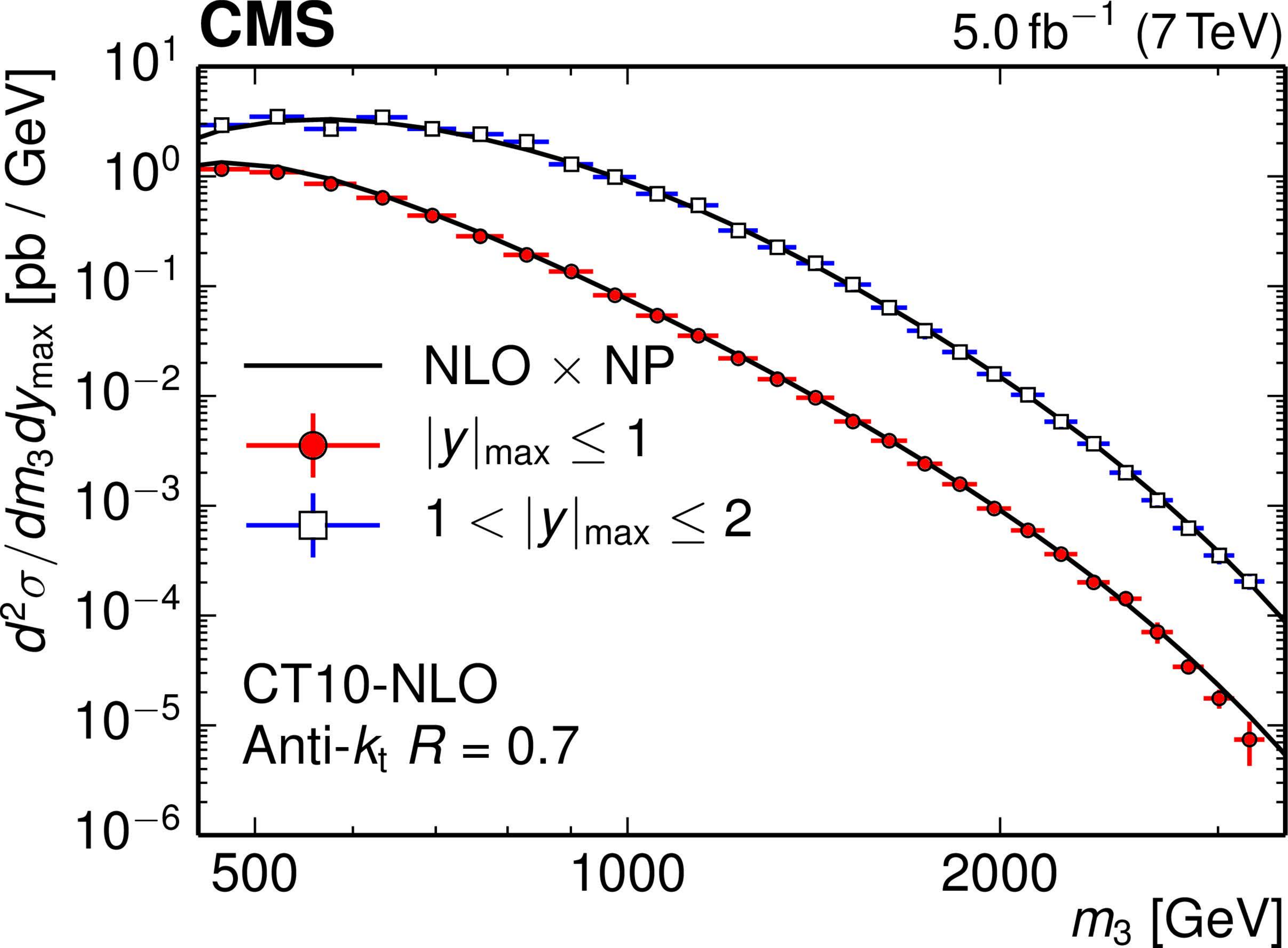}}
\mbox{\includegraphics[width=0.5\textwidth]{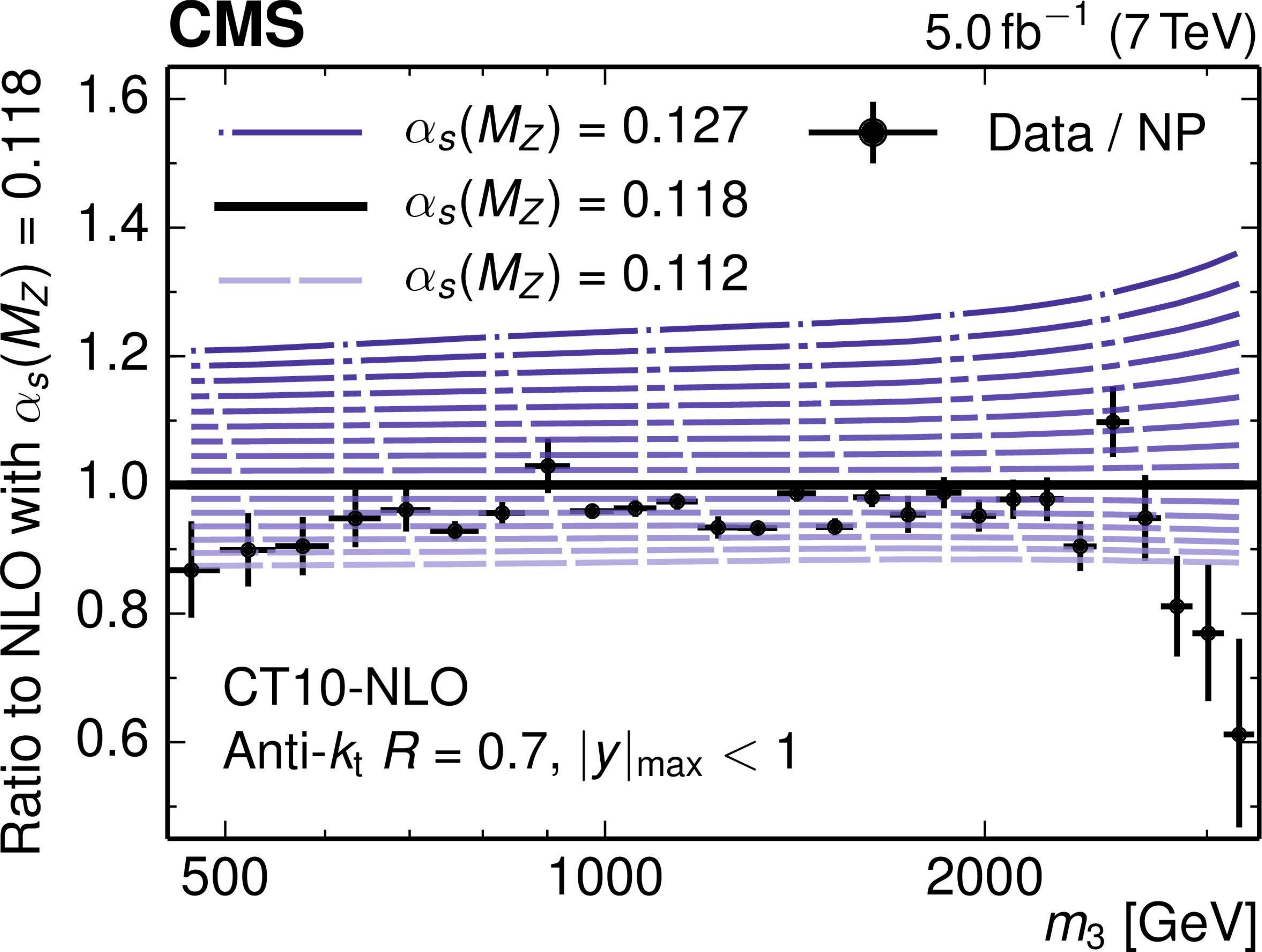}}
}
\caption{{\bf Left:} The CMS 3-jet mass cross section, as a function of 3-jet mass $m_3$, compared with
  the theory prediction for two regions in rapidity. The vertical
  error bars embody the total experimental uncertainty, whereas the
  horizontal error bars denote the bin widths~\cite{CMS:2014mna}. {\bf
    Right:} Ratio of the CMS measured 3-jet cross section in the
  central rapidity region, divided by the nonperturbative (NP)
  correction, to the theory prediction at next-to-leading order
  (NLO).~\cite{CMS:2014mna}.}
\label{fig:cms_3j}
\end{figure}

The doubly differential 3-jet mass cross section, expressed as a
function of the invariant mass $m_3$ (where $m_3^2 \equiv (p_1 + p_2 +
p_3)^2$) and maximum rapidity ($y_{\rm max}$) of the 3-jet system, is
proportional to $\alpha_{\mathrm{s}}^3$. Both the ATLAS~\cite{Aad:2014rma} and
CMS~\cite{CMS:2014mna} collaborations have measured this
observable. Figure~\ref{fig:cms_3j} (Left) depicts the CMS measured
differential cross section, which is described well by NLO
perturbative QCD (with nonperturbative corrections) over five orders
of magnitude and for 3-jet masses up to 3~TeV. In
Fig.~\ref{fig:cms_3j} (Right), a ratio of the data, corrected for
nonperturbative effects, to the NLO theory prediction using a nominal
value of $\alpha_{\mathrm{s}}(M_Z)=0.118$, is presented. In addition, theoretical
predictions involving $\alpha_{\mathrm{s}}(M_Z)$ values ranging from 0.112 to
0.127 in steps of 0.001 are shown. The extracted fitted result was
found to be~\cite{CMS:2014mna}
\begin{equation}
\alpha_{\mathrm{s}}(M_Z) = 0.1171\pm 0.0013({\rm exp})\pm 0.0024({\rm PDF})\pm 0.0008({\rm NP})\phantom{}^{+0.0069}_{-0.0040}({\rm scale}),
\end{equation}
with a total uncertainty ranging between 4\% and 6\%, again dominated
by $\mu_r = \mu_f \equiv m_3/2$.  To examine the running of $\alpha_{\mathrm{s}}$
with $Q$, this analysis was broken down into seven sub-datasets,
providing $\alpha_{\mathrm{s}}(Q)$ quantities at seven different values of
momentum transfer, including a highest value of 1402~GeV, which
constitutes the highest $Q$ value of $\alpha_{\mathrm{s}}$ measurements performed
to date~\cite{CMS:2014mna}.

\section{Top-Quark Pair Production Cross Section}

\begin{figure}
\centerline{
\mbox{\includegraphics[width=0.5\textwidth]{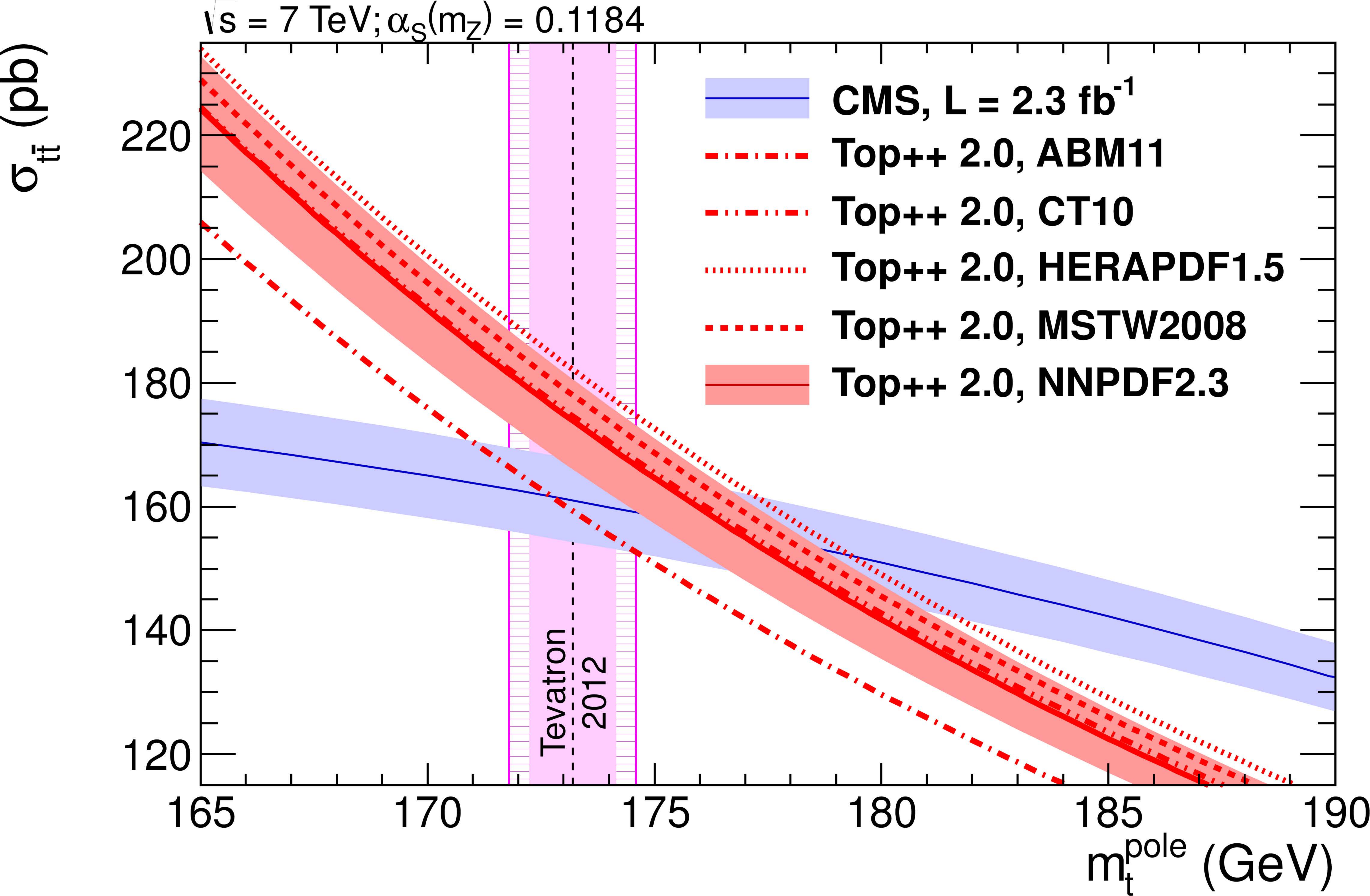}}
\mbox{\includegraphics[width=0.5\textwidth]{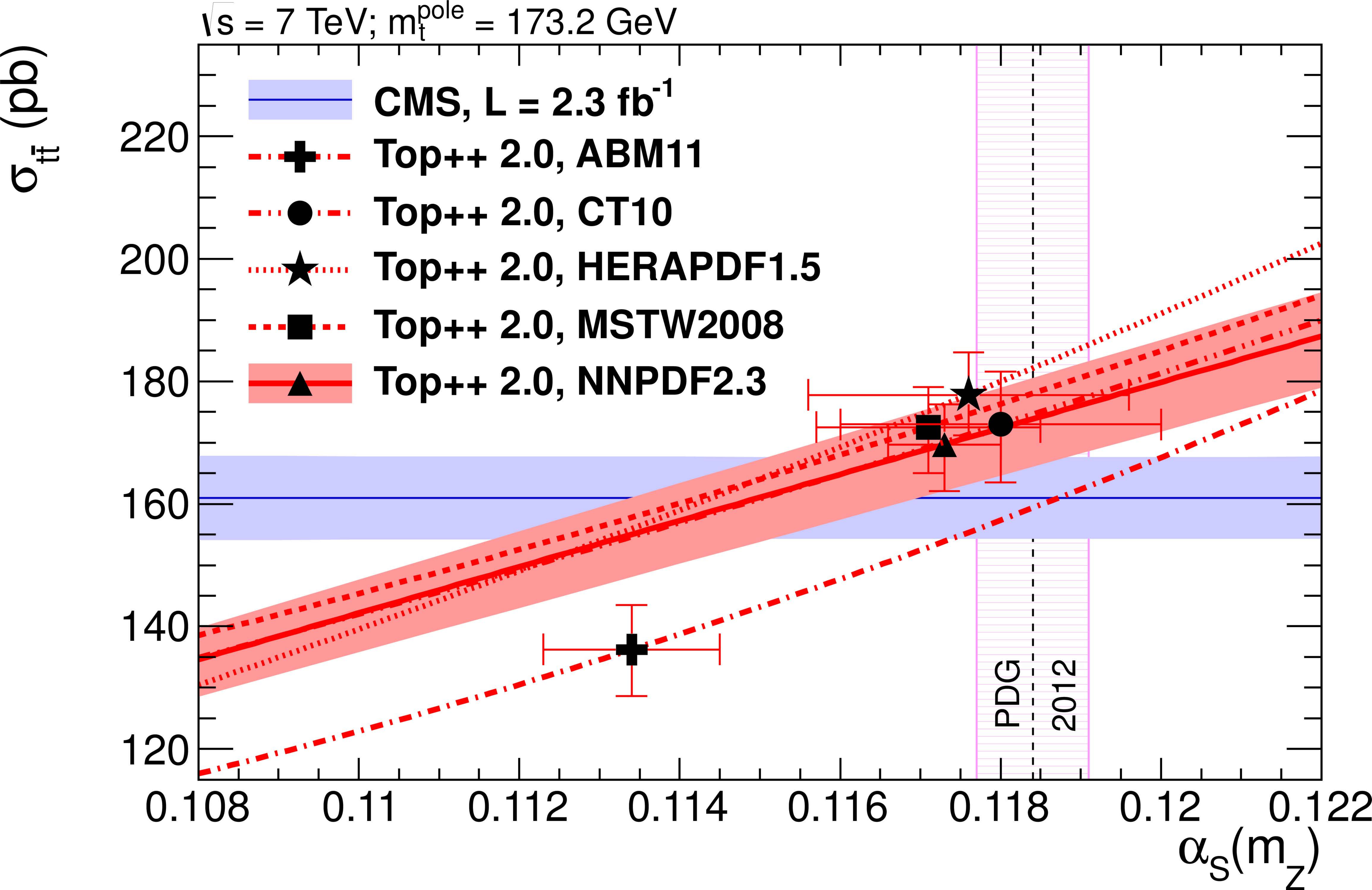}}
}
\caption{Predicted $t\bar{t}$ cross section at NNLO+NNLL as a function
  of the top-quark pole mass ({\bf Left}) and of the strong coupling
  constant $\alpha_{\mathrm{s}}(M_Z)$ ({\bf Right}), using five different NNLO
  PDF sets, compared to the cross section measured by
  CMS~\cite{Chatrchyan:2012bra} assuming $m_t = m_t^{\rm
    pole}$~\cite{Chatrchyan:2013haa}.}
\label{fig:cms_ttbar_constrained}
\end{figure}

A first measurement of $\alpha_{\mathrm{s}}$ in a study of top-quark production
has been achieved by the CMS collaboration~\cite{Chatrchyan:2013haa}.
The inclusive top-antitop ($t\bar{t}$)
production cross section, $\sigma_{t\bar{t}}$, in 2.3~fb$^{-1}$ of
$pp$ collisions at a centre-of-mass energy of $\sqrt{s} =
7$~TeV~\cite{Chatrchyan:2012bra} is used for the measurement.
To accomplish this, a theoretical
prediction involving QCD production at next-to-next-to-leading-order
(NNLO) precision combined with soft-gluon resummation at
next-to-next-to-leading-log (NNLL) precision~\cite{Czakon:2013goa} was compared with the
experimental $\sigma_{t\bar{t}}$ for the assumed scales $\mu_r$ and
$\mu_f$ set to the top-quark pole mass, $m_t^{\rm pole}$.
In $\sigma_{t\bar{t}}$ cross-section calculations, $\alpha_{\mathrm{s}}$ appears both
in the expression for the parton-parton interactions and in the QCD evolution of the proton PDFs.
Figure~\ref{fig:cms_ttbar_constrained} (Left) shows the
predicted $\sigma_{t\bar{t}}$ for several NNLO PDF sets as a function
of $m_t^{\rm pole}$ and with $\alpha_{\mathrm{s}}$ constrained to a value of
$0.1184\pm
0.0007$~\cite{Beringer:1900zz}. Figure~\ref{fig:cms_ttbar_constrained}
(Right) depicts the predicted $\sigma_{t\bar{t}}$ as a function of
$\alpha_{\mathrm{s}}(M_Z)$, assuming $m_t^{\rm pole} = 173.2\pm 1.4$~GeV.

For a given NNLO PDF set, identifying the intersection of the
$\sigma_{t\bar{t}}$ prediction with the CMS measured value in
Fig.~\ref{fig:cms_ttbar_constrained} (Right) yields a precise
determination of $\alpha_{\mathrm{s}}(M_Z)$.  Using this technique for the
centrally treated NNPDF2.3 PDF, the CMS collaboration obtained the
most precise measurement of $\alpha_{\mathrm{s}}$ at a hadron collider, with a
total uncertainty of 2.4\%~\cite{Chatrchyan:2013haa}:
\begin{equation}
\alpha_{\mathrm{s}}(M_Z) = 0.1151\phantom{}^{+0.0017}_{-0.0018} ({\rm exp})\pm
\phantom{}^{+0.0013}_{-0.0011}({\rm PDF})\pm 0.0013(m_t^{\rm pole})
\pm 0.0008(E_{\rm LHC})\phantom{}^{+0.0009}_{-0.0008}({\rm scale}).
\end{equation}
Unlike for the earlier described $\alpha_{\mathrm{s}}$ measurements, the
$\pm46$~GeV uncertainty in the 7~TeV $pp$ collision energy, $E_{\rm
  LHC}$, manifests as a sizeable source of uncertainty on
$\alpha_{\mathrm{s}}(M_Z)$ using the $\sigma_{t\bar{t}}$ technique, tantamount
to that due to scale-related sources.

\begin{figure}
\centering
\includegraphics[height=2.3in]{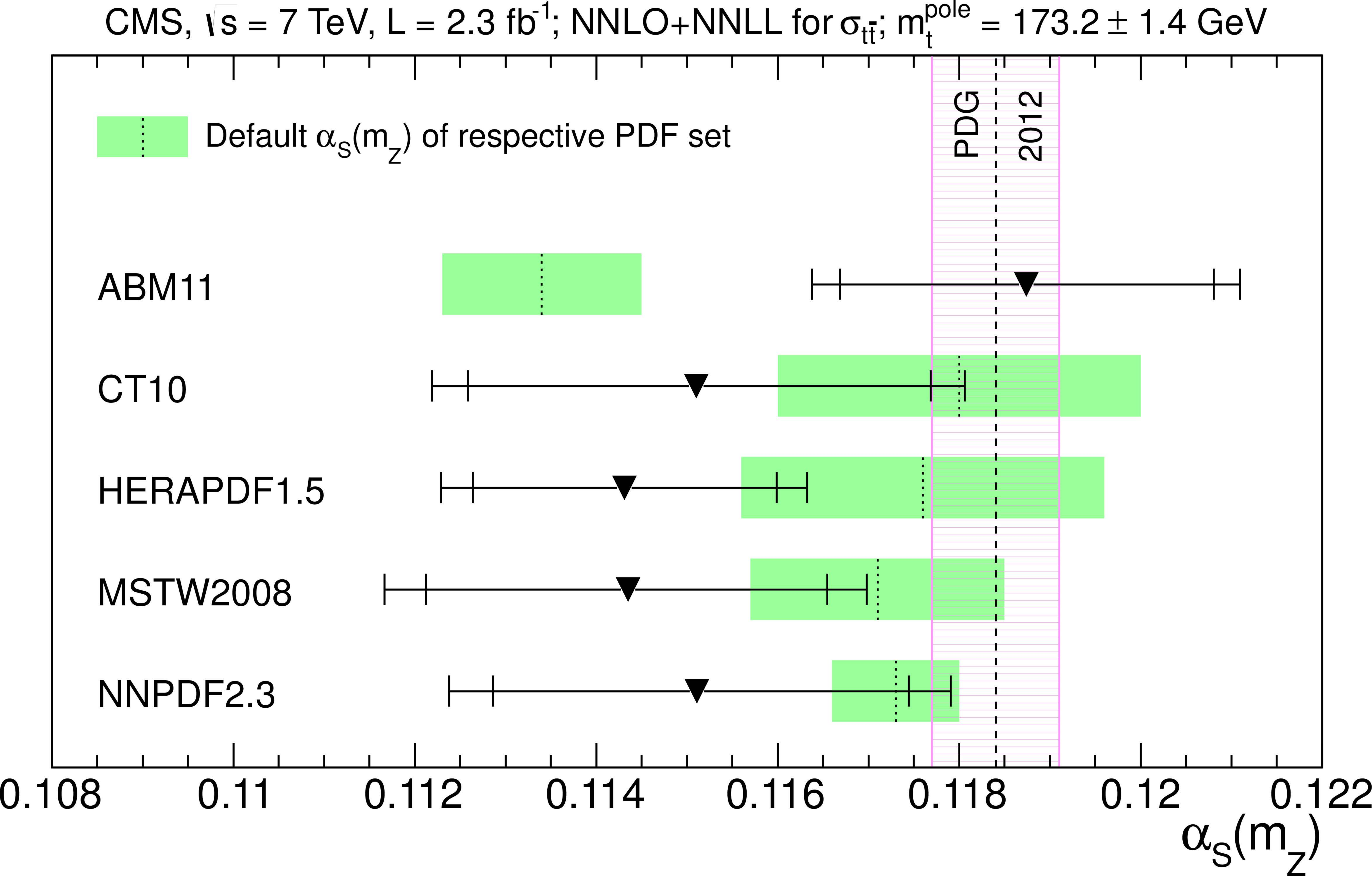}
\caption{CMS results~\cite{Chatrchyan:2013haa} obtained for
  $\alpha_{\mathrm{s}}(M_Z)$ from the measured $t\bar{t}$ cross
  section~\cite{Chatrchyan:2012bra} together with the prediction at
  NNLO+NNLL using different NNLO PDF sets.  The inner error bars
  include the uncertainties on the measured cross section and on the
  LHC beam energy as well as the PDF and scale uncertainties on the
  predicted cross section. The outer error bars additionally account
  for the uncertainty on $m_t^{\rm pole}$. For comparison, the
  $\alpha_{\mathrm{s}}(M_Z)$ world average~\cite{Beringer:1900zz} with its
  uncertainty is shown as a hatched band. For each PDF set, the
  default $\alpha_{\mathrm{s}}(M_Z)$ value, used natively by that PDF set, and
  its uncertainty are indicated using a dotted line and a shaded
  band.}
\label{fig:cms_ttbar_alphaS_results}
\end{figure}

Figure~\ref{fig:cms_ttbar_alphaS_results} depicts results for
$\alpha_{\mathrm{s}}(M_Z)$ determined using various different NNLO PDF sets. The
NNPDF2.3, CT10, MSTW2008, and HERAPDF1.5 PDF sets are somewhat lower
than the world average~\cite{Beringer:1900zz}, though generally still
compatible. While the ABM11 result for $\alpha_{\mathrm{s}}$ is compatible with
the world average, its default natively assumed
value is seen to be markedly different.

\section{Conclusion}

\begin{figure}
\centering
\includegraphics[height=2.6in]{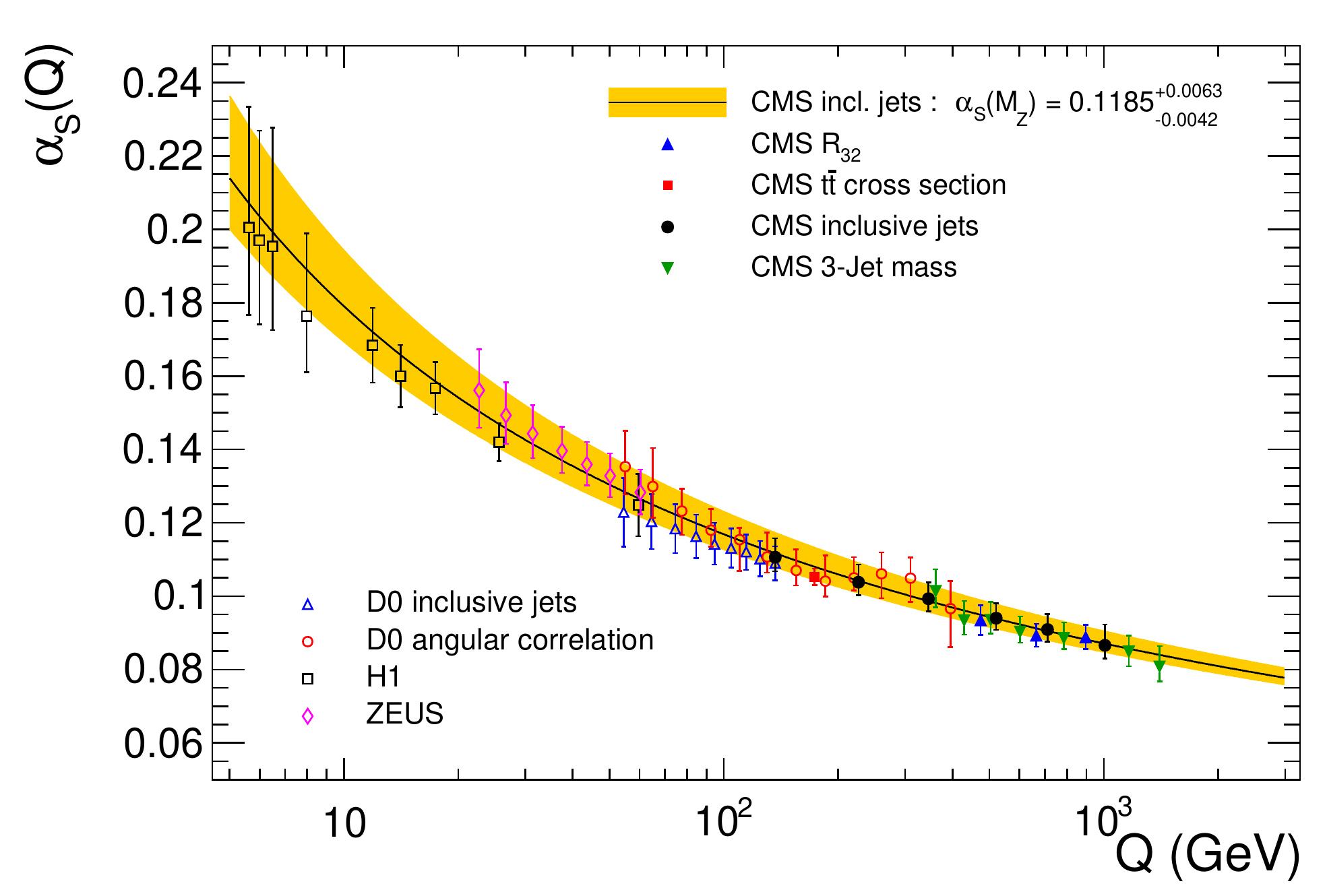}
\caption{The strong coupling $\alpha_{\mathrm{s}}(Q)$ (solid line)
  and its total uncertainty (band) as determined in the
  Ref.~\cite{Khachatryan:2014waa} analysis using a 2-loop solution to
  the RGE as a function of the momentum transfer $Q=p_{{\mathrm{T}}}$ and 7~TeV
  inclusive jet cross section measurements described in
  Ref.~\cite{Chatrchyan:2012bja}. Extractions of $\alpha_{\mathrm{s}}(Q)$ in six
  separate ranges of $Q$ are shown together with results from the
  H1~\cite{Aaron:2009vs,Aaron:2010ac}, ZEUS~\cite{Abramowicz:2012jz},
  and D0~\cite{Abazov:2009nc,Abazov:2012lua} experiments at the HERA
  and Tevatron colliders.  Additional recent CMS measurements are also
  included~\cite{Chatrchyan:2013txa,Chatrchyan:2013haa,CMS:2014mna}.}
\label{fig:summary_plot_running}
\end{figure}

Figure~\ref{fig:summary_plot_running}~\cite{CMS:summary_plots} places
the discussed CMS $\alpha_{\mathrm{s}}(Q)$ results alongside some select
measurements from electron-proton and proton-antiproton colliders,
visually demonstrating the mutual consistency in the running of the
$\alpha_{\mathrm{s}}$ coupling with scale $Q$, as well as consistency with the
QCD RGE prediction described in Ref.~\cite{Khachatryan:2014waa}. Note
the comparatively small uncertainty on the (red square) point showing
the $\alpha_{\mathrm{s}}$ value derived using the $\sigma_{t\bar{t}}$
technique~\cite{Chatrchyan:2013haa}, the first NNLO measurement of $\alpha_{\mathrm{s}}$
performed at a hadron collider.  To date, no significant deviations
from the QCD RGE running have been observed, and
first explorations of $\alpha_{\mathrm{s}}$ running into the TeV regime have commenced.

\begin{figure}
\centering
\includegraphics[height=3.2in]{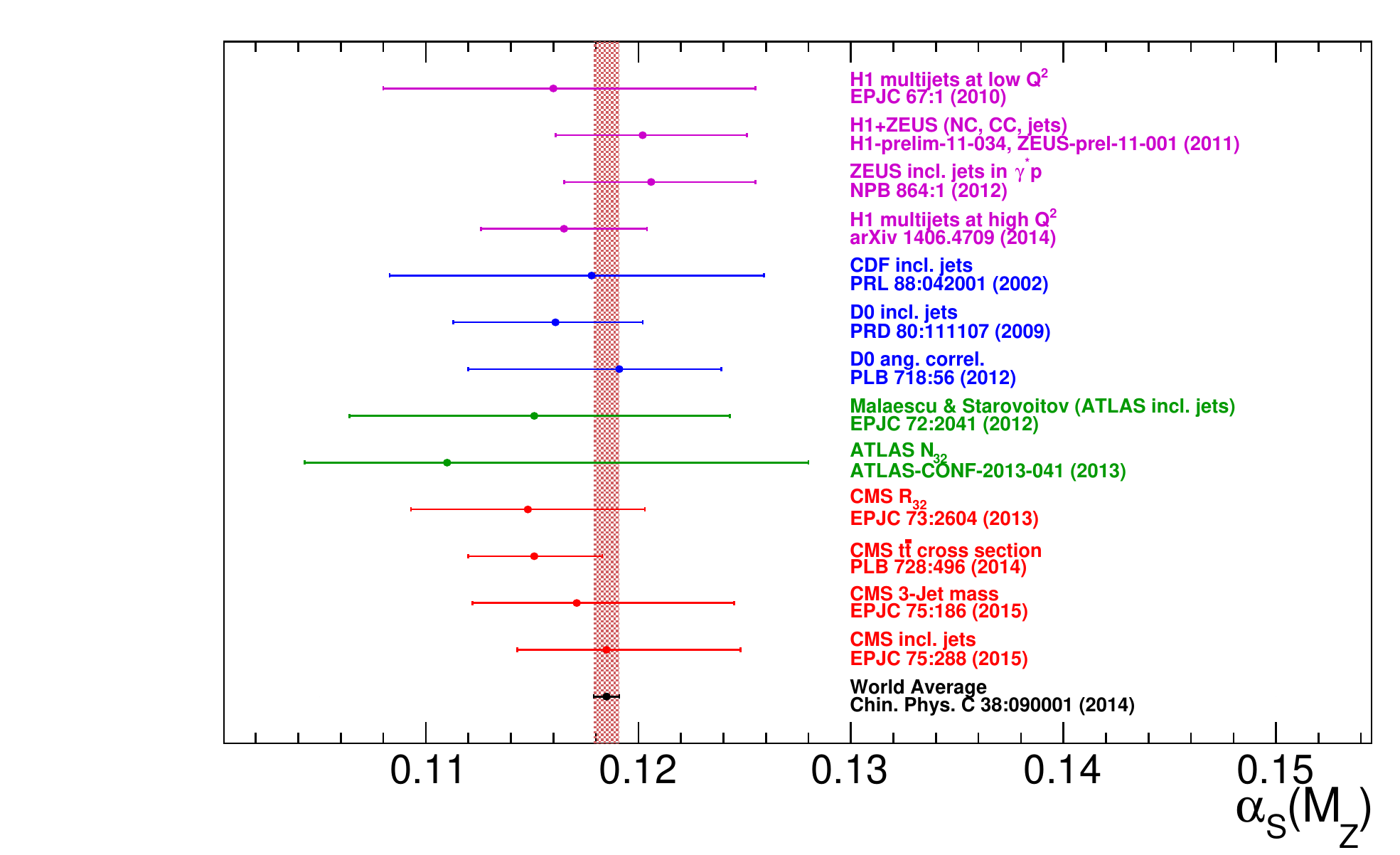}
\caption{An overview of recent $\alpha_{\mathrm{s}}(M_Z)$ measurements made
  in hadron collisions (with citations), and the current world-average
  value~\cite{Agashe:2014kda}.}
\label{fig:summary_plot_alphasMZ}
\end{figure} 

Figure~\ref{fig:summary_plot_alphasMZ}~\cite{CMS:summary_plots}
provides an overview of the $\alpha_{\mathrm{s}}(M_Z)$ measurements described
here, as well as some earlier results from colliders involving
hadrons. The indicated measurements agree with each other and with the
current world average~\cite{Agashe:2014kda}. New
7~TeV~\cite{ATLAS:2015yaa} and 8~TeV results are on the way from the
LHC, with others anticipated at 13~TeV and 14~TeV. In the interim,
theoretical improvements at the NNLO level will need to be provided to
enable significant further precision increases in the future.

\Acknowledgements I am grateful for support from the Natural Sciences
and Engineering Research Council (NSERC) of Canada and the Fonds de
Recherche du Qu\'ebec --- Nature et Technologies (FRQNT).

\end{document}